\documentclass[sigconf]{acmart}

\newcommand{\dcimgwidth}{0.87\textwidth}
\usepackage{algorithm}
\usepackage{algpseudocode}
\usepackage{tikz}
\newcommand{\inversecircledsmall}[1]{%
  \ensuremath{%
    \mathord{%
      \tikz[baseline=(char.base)]{
        \node[
          shape=circle,
          draw=black,
          fill=black,
          inner sep=0.8pt,
          outer sep=0pt
        ] (char) {\small\textcolor{white}{\sffamily\bfseries #1}};
      }
    }%
  }%
}
\AtBeginDocument{%
  }

\copyrightyear{2026}
\acmYear{2026}
\setcopyright{cc}
\setcctype{by}
\acmConference[ICCAD '26]{IEEE/ACM International Conference on Computer-Aided Design}{November 08--12, 2026}{San Jose, CA, USA}
\acmBooktitle{IEEE/ACM International Conference on Computer-Aided Design (ICCAD '26), November 08--12, 2026, San Jose, CA, USA}
\acmDOI{10.1145/3831252.3834102}
\acmISBN{979-8-4007-2873-0/2026/11}

\begin{document}

\title{APT: Accelerating Diffusion Transformers via Attention Probability-Guided Pruning and Quantization}

\author{Sungyeob Yoo}
\affiliation{%
  \institution{KAIST}
  \city{Daejeon}
  \country{Republic of Korea}
}
\email{sungyeob.yoo@kaist.ac.kr}

\author{Seeyeon Kim}
\affiliation{%
  \institution{KAIST}
  \city{Daejeon}
  \country{Republic of Korea}}
\email{seeyakim@kaist.ac.kr}

\author{Joonyong Park}
\affiliation{%
  \institution{KAIST}
  \city{Daejeon}
  \country{Republic of Korea}}
\email{wndy2001@kaist.ac.kr}

\author{Seunghee Han}
\affiliation{%
  \institution{KAIST}
  \city{Daejeon}
  \country{Republic of Korea}}
\email{shhan1755@kaist.ac.kr}

\author{Joo-Young Kim}
\affiliation{%
  \institution{KAIST}
  \city{Daejeon}
  \country{Republic of Korea}}
\email{jooyoung1203@kaist.ac.kr}


\begin{abstract}
Recent advances in generative AI have significantly increased the demand for high-resolution image and video generation, positioning diffusion models as a core technology.
Among them, Diffusion Transformers (DiTs) have emerged as the state-of-the-art (SOTA) models due to their scalability and output quality.
However, self-attention in DiTs incurs significant computational overhead, leading to excessively long latency as the complexity grows with the fourth power of the output resolution.
While prior works have attempted to mitigate this cost using sparsity and quantization techniques, they fall short of effectively reducing the computational cost in high-resolution DiTs.

In this paper, we present APT, a software--hardware co-designed accelerator for high-resolution DiTs.
APT leverages attention probabilities as a unified importance metric to jointly optimize computation through fine-grained pruning and adaptive precision scaling.
At the algorithm level, we propose Attention Probability-guided Adaptive Dual Thresholding (APDT), which dynamically performs element selection and precision assignment using dual thresholds.
To ensure compatibility with memory-efficient FlashAttention, we introduce Timestep-Aware FlashAttention (TAFA), which predicts attention probabilities across timesteps by exploiting temporal similarity.
At the architecture level, we co-design a specialized accelerator that efficiently supports irregular sparsity and dual-precision execution, featuring dynamic mask management, address translation, dual-precision compute units, and a tile-based dataflow.
Finally, we evaluate APT on SOTA DiT models, including PixArt-$\alpha$, Stable Diffusion 3, and FLUX.
APT achieves up to 8.16$\times$ speedup and 14.98$\times$ higher energy efficiency over NVIDIA A100, and up to 3.01$\times$ speedup and 2.04$\times$ higher energy efficiency over EXION, a SOTA diffusion model accelerator.
\end{abstract}



\maketitle

\section{Introduction}
\label{sec:sec1}

The recent surge in generative AI has positioned diffusion models as the state-of-the-art (SOTA) backbone for high-fidelity image and video synthesis~\cite{10.5555/3495724.3496298, rombach2021highresolution, 10.5555/3586589.3586636, 9887996, pmlr-v162-nichol22a}.
Diffusion Transformer (DiT) models such as PixArt-$\alpha$~\cite{chen2023pixartalpha}, Stable Diffusion 3 (SD3)~\cite{10.5555/3692070.3692573}, and FLUX~\cite{flux2024} replace U-Nets with Transformers~\cite{10.5555/3295222.3295349} and have become the standard for high-quality generation.

\begin{figure}[t]
    \centering
    \includegraphics[width=0.44\textwidth]{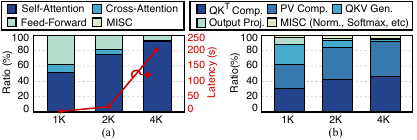}
    \vspace{-0.15in}
    \caption{(a) Computation breakdown and latency measured on NVIDIA A100 (b) Operation breakdown in self-attention}
    \vspace{-0.1in}
    \label{fig:fig1}
\end{figure}

Despite their flexibility, DiT models suffer from significant computational overhead as resolution increases.
As shown in Figure~\ref{fig:fig1}(a), self-attention accounts for over 90\% of total computation at 4K resolution, making it the dominant bottleneck.
This overhead grows rapidly with resolution.
Since the token count grows as $N^2$ for an image of resolution $N$, and self-attention scales quadratically with token count, the two core operations---$query$$\times$$key^\top$ ($QK^\top$) and $attention\_probability$$\times$$value$ ($PV$)---yield $\mathcal{O}(N^4)$ overall complexity.
Figure~\ref{fig:fig1}(b) confirms that $QK^\top$ and $PV$ dominate self-attention, underscoring the necessity of reducing their cost.
To address this, pruning and quantization have been explored as two widely studied techniques to reduce computation~\cite{exion, cho2024characterization, he2023ptqd, Shang2023ptqd, zhao2025viditq, chen2024qdit}.
While pruning and quantization have usually been applied separately, both aim to reduce computation on less essential elements: pruning eliminates them entirely, while quantization reduces their precision.
Here, the key challenge lies in identifying which elements are essential and how to adapt these decisions across timesteps, so that both techniques can be applied jointly through a unified metric.

We observe that attention probabilities provide a unified importance metric for jointly guiding both techniques.
After Softmax, they concentrate on a few dominant entries while suppressing the rest, revealing which elements can be pruned and which require reduced precision.
Moreover, this distribution remains highly similar across adjacent timesteps in DiT, so importance information from one timestep can inform decisions at the next.
However, while this similarity holds between adjacent steps, the overall distribution evolves as denoising progresses, shifting from nearly uniform to sharply concentrated, so decisions must adapt accordingly.
Translating these observations into a practical system poses two additional challenges.
First, modern DiT pipelines rely on FlashAttention, which avoids materializing attention probabilities to save memory.
Second, the element-wise sparsity and per-element precision assignment from probability-guided decisions create irregular patterns that existing hardware cannot efficiently handle.

Motivated by this, we present APT, a software--hardware co-designed accelerator for high-resolution DiTs.
At the algorithm level, we propose Attention Probability-guided Adaptive Dual Thresholding (APDT), which applies two coordinated thresholds---one for pruning and one for precision assignment---to classify each attention element into skipped, low-precision, or high-precision execution based on attention probability statistics.
To enable APDT within FlashAttention, we introduce Timestep-Aware FlashAttention (TAFA), which predicts attention probabilities by reusing normalization statistics from adjacent timesteps.
At the architecture level, we design a specialized architecture featuring tile-based dataflow, dynamic mask management, address translation, and dual-precision compute units to efficiently handle the resulting irregular sparsity.
Figure~\ref{fig:fig2} provides an overview.
Our contributions are as follows:

\begin{figure}[t]
    \centering
    \includegraphics[width=0.44\textwidth]{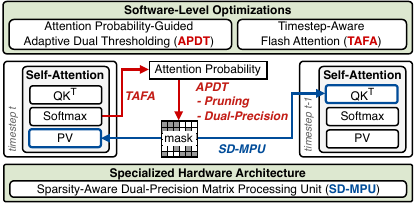}
    \vspace{-0.15in}
    \caption{Overview of APT}
    \vspace{-0.2in}
    \label{fig:fig2}
\end{figure}

\begin{itemize}
\item We propose APT, a software--hardware co-designed accelerator for high-resolution DiTs that jointly leverages pruning and quantization guided by attention probabilities.
\item We introduce APDT for adaptive element-wise pruning and dual-precision quantization, and TAFA for enabling APDT within FlashAttention's memory-efficient framework.
\item APT features a specialized hardware architecture with fine-grained sparsity support, dual-precision execution, and efficient mask handling via tile-based dataflow.
\item APT achieves up to 8.16$\times$ and 3.01$\times$ speedup, and up to 14.98$\times$ and 2.04$\times$ higher energy efficiency over the A100 and EXION, respectively.
\end{itemize}

\section{Background and Motivation}
\label{sec:sec2}

\subsection{Overview of Diffusion Transformer}
\label{sec:sec2.1}

\begin{figure}[t]
    \centering
    \includegraphics[width=0.44\textwidth]{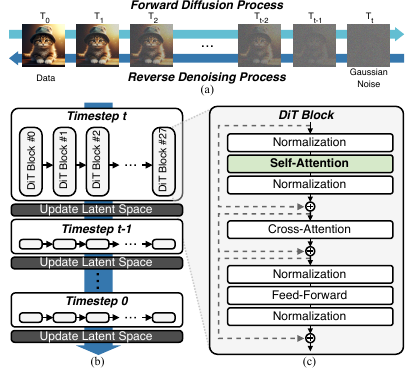}
    \vspace{-0.13in}
    \caption{(a) Overview of diffusion process (b) Execution flow of DiT (c) Architecture of DiT block}
    \vspace{-0.15in}
    \label{fig:fig3}
\end{figure}

The diffusion model is a generative model that synthesizes data by gradually denoising Gaussian noise.
As shown in Figure~\ref{fig:fig3}(a), it removes noise step by step through a trained denoising process to produce high-quality outputs.
Recently, SOTA diffusion models such as PixArt-$\alpha$~\cite{chen2023pixartalpha}, SD3~\cite{10.5555/3692070.3692573}, and FLUX~\cite{flux2024} have adopted Transformer-based backbones, known as Diffusion Transformers (DiTs).
DiT progressively refines the latent representation across timesteps using a sequence of Transformer blocks, each consisting of self-attention, cross-attention, feed-forward layers, and normalization, as shown in Figure~\ref{fig:fig3}(b) and (c).
Among these components, self-attention is the dominant operation.
As shown in Figure~\ref{fig:fig4}(a), self-attention first generates query, key, and value matrices, then computes scaled attention scores via $QK^\top$, applies Softmax to produce attention probabilities, aggregates context through $PV$, and projects the result to the output.

\subsection{Temporal Similarity in DiT}
\label{sec:sec2.2}

\begin{figure}[t]
    \centering
    \includegraphics[width=0.44\textwidth]{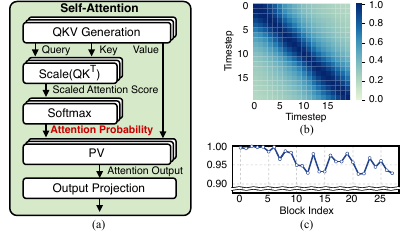}
    \vspace{-0.13in}
    \caption{(a) Computational flow of self-attention (b) Cosine similarity of attention probability between timesteps (c) Cosine similarity between consecutive timesteps across blocks}
    \vspace{-0.15in}
    \label{fig:fig4}
\end{figure}

DiT models exhibit temporal similarity in attention maps~\cite{ditto}.
Unlike standard Transformers, DiT models process inputs that evolve gradually across timesteps due to the iterative nature of the diffusion process, resulting in high similarity in attention probability distributions between adjacent steps.
As shown in Figure~\ref{fig:fig4}(b) and (c), the cosine similarity of attention probabilities exceeds 0.92 for consecutive steps in most blocks, and decays gradually with larger temporal gaps.
This temporal similarity suggests that attention probability information from one timestep can be reused in adjacent steps to reduce redundant computation.
Specifically, if the attention probability distribution is known from the previous timestep ($t{+}1$), it can serve as a proxy for identifying which elements are important at the current step, enabling both pruning decisions and precision assignment without recomputing full attention scores.
This property is unique to diffusion models and does not hold for standard Transformer inference.
Note that experimental results up to Section~\ref{sec:sec3} are based on PixArt-$\alpha$.

\subsection{Limitations of Prior Approaches}
\label{sec:sec2.3}

\textbf{Softmax pruning}~\cite{lu2021sanger, cho2024characterization} exploits the fact that Softmax amplifies large attention scores while suppressing small ones, resulting in attention probabilities where only a few entries have significant magnitude.
By thresholding these probabilities, low-importance elements can be removed from $PV$ computation with minimal accuracy loss.
However, it faces three limitations in DiTs: (1) a fixed threshold ignores distributional variance across timesteps, which can either remove important information or miss available sparsity, (2) mask reuse degrades as temporal similarity weakens over time, and (3) storing full binary masks at high resolutions requires prohibitive memory (448\,GiB at 4K), far exceeding GPU capacity.

\textbf{Quantization} for DiTs~\cite{wu2024ptq4dit, li2024svdquant} has shown that linear and convolutional layers can operate at low precision.
However, SVDQuant~\cite{li2024svdquant} leaves $QK^\top$ and $PV$ in full precision, as quantizing these activation--activation operations results in substantial accuracy loss. Since these operations dominate the cost at high resolutions, quantization alone provides limited benefit for the attention bottleneck.

\textbf{FlashAttention (FA)}~\cite{dao2023flashattention2} is a memory-efficient self-attention algorithm that avoids materializing the full attention probability matrix by reordering computations into tiles.
FA loads the query matrix into SRAM and processes key and value matrices tile by tile, performing partial Softmax per tile and accumulating scaling factors to reconstruct the final output.
Specifically, for each query tile $i$ and key-value tile $j$, FA computes tile-level unnormalized exponentials $\mathbf{P}_i^{\prime(j)}$ and accumulates exponential sums $\mathbf{l}_i^{(j)}$ across tiles, but the final Softmax scaling factor $\mathbf{R}_i^{(j)}$ that normalizes each tile's contribution is only available after all tiles are processed.
This design prevents direct access to attention probabilities, making FA incompatible with Softmax pruning and other attention-probability-based optimizations.
Therefore, a new prediction mechanism is required that approximates attention probabilities without compromising FA's memory efficiency.

\section{APT's Algorithm Optimizations}
\label{sec:sec3}

APT introduces two algorithm-level optimizations.
(1) Attention Probability-guided Adaptive Dual Thresholding (APDT) jointly applies pruning and dual-precision quantization at element-wise granularity.
(2) Timestep-Aware FlashAttention (TAFA) enables APDT within FA by combining normalization statistics from the previous timestep with the current timestep's partial sums to predict attention probabilities without full probability materialization.

\begin{figure}[t]
    \centering
    \includegraphics[width=0.44\textwidth]{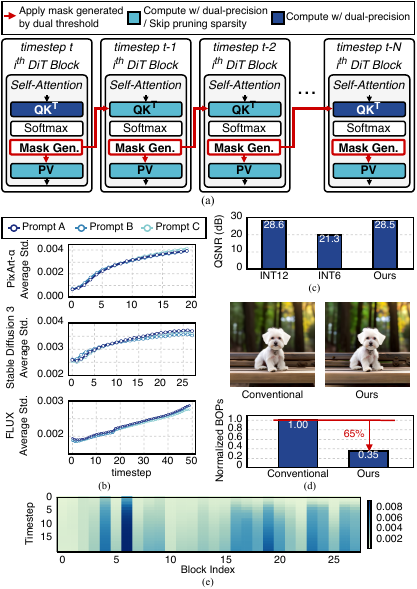}
    \vspace{-0.13in}
    \caption{(a) Execution flow of APDT across timesteps (b) Average standard deviation of attention probabilities over timesteps (c) Comparison of accuracy across precision settings (d) Comparison of normalized computation under similar output quality (e) Per-block and per-timestep standard deviation of attention probabilities}
    \vspace{-0.15in}
    \label{fig:fig6}
\end{figure}

\subsection{Attention Probability-Guided Adaptive Dual Thresholding}
\label{sec:sec3.1}

APDT reduces self-attention cost by leveraging the standard deviation of attention probabilities as a metric to adaptively guide pruning and precision assignment.
As shown in Figure~\ref{fig:fig6}(a), two thresholds are applied at each timestep: the pruning threshold removes unimportant elements, while the quantization threshold assigns precision.
Elements above the quantization threshold use 12-bit precision; elements between the two thresholds use 6-bit; elements below the pruning threshold are skipped entirely.

A key observation is that the standard deviation of attention probabilities increases as denoising progresses (Figure~\ref{fig:fig6}(b)): early in the process, the latent input is highly noisy and attention scores are distributed uniformly, but in later stages, attention becomes concentrated as semantically meaningful regions emerge.
Moreover, the standard deviation differs substantially across blocks within the same timestep (Figure~\ref{fig:fig6}(e)), further motivating adaptive thresholding.
APDT adjusts thresholds per head and timestep:
$$
threshold_{h}(t) = \alpha \cdot \frac{\sigma_{h}(t) - \sigma_{h,\min}}{\sigma_{h,\max} - \sigma_{h,\min}} + \beta
$$
where $\sigma_{h}(t)$ is the profiled standard deviation of attention head $h$ at timestep $t$, obtained through a one-time offline profiling step per model.
The same formula is applied twice with separate $(\alpha, \beta)$ pairs for the pruning and quantization thresholds, respectively.
The quantization threshold is kept strictly above the pruning threshold by assigning larger $\alpha$ and $\beta$ values to the quantization pair.
The hyperparameters $\alpha$ and $\beta$ are determined per model, and $\sigma_{h,\min}$, $\sigma_{h,\max}$ denote the minimum and maximum of $\sigma_h(t)$ observed across all timesteps during offline profiling.
With these $\sigma$-adaptive thresholds, APDT preserves more information during the early timesteps when attention is distributed uniformly, while allowing greater computational reduction in later stages when attention concentrates on fewer elements.

APDT generates high- and low-precision execution masks that are reused from the previous timestep ($t{+}1$) via temporal similarity (Figure~\ref{fig:fig4}(c)), avoiding redundant $QK^\top$ computation.
Pruning and precision decisions are applied independently to individual attention probability matrix elements, resulting in element-wise dynamic sparsity.
Masks are generated at runtime since attention distributions are input-dependent.
To prevent quality degradation, APDT employs periodic refresh (every 7--10 steps), which reintroduces all pruned elements into the low-precision mask, and $\sigma$-driven refresh per head, triggered independently when the offline-profiled head-wise $\sigma$ drops by more than 5\%--10\% of its observed range.
Quantization uses symmetric uniform quantization at the tile level.
As shown in Figure~\ref{fig:fig6}(c)--(d), APDT achieves a quantization signal-to-noise ratio (QSNR)~\cite{10.1145/3579371.3589351} of 28.5\,dB, comparable to INT12 (28.6\,dB), and reduces BOPs by up to 65\% versus fixed-threshold pruning.

\begin{figure}[t]
    \centering
    \includegraphics[width=0.44\textwidth]{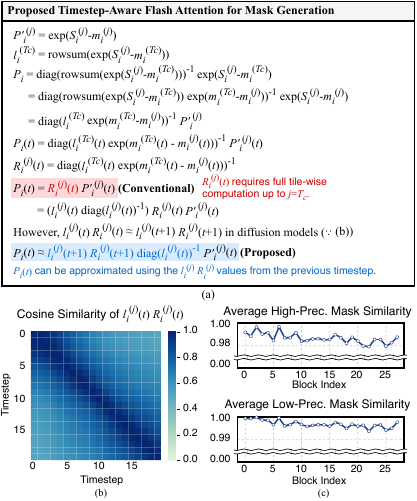}
    \vspace{-0.17in}
    \caption{(a) Process of TAFA (b) Cosine similarity of $\mathbf{l}_i^{(j)} \cdot \mathbf{R}_i^{(j)}$ between timesteps (c) Comparison of high- and low-precision execution masks generated from TAFA and full Softmax}
    \vspace{-0.15in}
    \label{fig:fig7}
\end{figure}

\subsection{Timestep-Aware FlashAttention}
\label{sec:sec3.2}

As described in Section~\ref{sec:sec2.3}, FA prevents direct access to attention probabilities within its tiled execution.
TAFA addresses this by reusing normalization statistics from the previous timestep to predict per-tile attention probabilities, as illustrated in Figure~\ref{fig:fig7}(a).
Recall that for the $i$-th query tile and $j$-th key-value tile (where $j$ ranges from 1 to $T_c$, the total number of key-value tiles), FA computes $\mathbf{S}_i^{(j)} = Q_i K_j^\top$ and the unnormalized exponential $\mathbf{P}_i^{\prime(j)} = \exp(\mathbf{S}_i^{(j)} - \mathbf{m}_i^{(j)})$, where $\mathbf{m}_i^{(j)}$ is the row-wise maximum.
The scaling factor $\mathbf{R}_i^{(j)}$ requires statistics from all $T_c$ tiles, but TAFA overcomes this by reusing $\mathbf{R}_i^{(j)}$ from timestep $t{+}1$:
\[
\mathbf{P}_i(t) \approx \underbrace{
\mathbf{l}_i^{(j)}(t{+}1) \cdot \mathbf{R}_i^{(j)}(t{+}1) \cdot \text{diag}(\mathbf{l}_i^{(j)}(t))^{-1}
}_{\text{TAFA factor}} \cdot \mathbf{P}_i^{\prime(j)}(t)
\]
The first term, $\mathbf{l}_i^{(j)}(t{+}1) \cdot \mathbf{R}_i^{(j)}(t{+}1)$, captures the ratio of each tile to the full attention probability at $t{+}1$; this ratio is stable across adjacent timesteps because attention distributions in DiT change gradually.
As shown in Figure~\ref{fig:fig7}(b), the cosine similarity of $\mathbf{l}_i^{(j)} \cdot \mathbf{R}_i^{(j)}$ between consecutive timesteps consistently exceeds 0.95.
Figure~\ref{fig:fig7}(c) confirms that execution masks from TAFA achieve $>$97\% similarity to those from full Softmax across all blocks.
By operating within FA's tiled framework, TAFA reduces on-chip memory requirements by up to 103.4$\times$ compared to naive APDT integration that materializes the full attention probability matrix at 4K resolution.
\section{APT's Hardware Architecture}
\label{sec:sec4}

\begin{figure*}[t]
    \centering
    \includegraphics[width=\dcimgwidth]{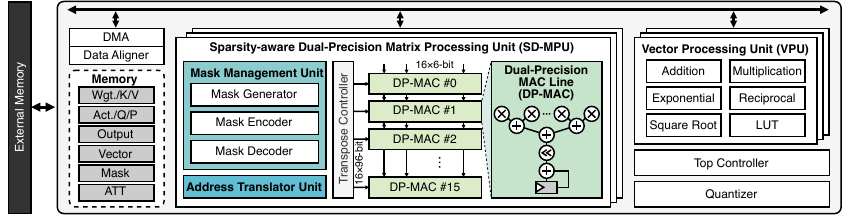}
    \vspace{-0.13in}
    \caption{Hardware architecture of APT}
    \vspace{-0.12in}
    \label{fig:fig8}
\end{figure*}

To exploit APDT's fine-grained dynamic sparsity and dual-precision execution, we propose APT, a specialized hardware architecture.
The core design challenges are: (1) generating and applying element-wise masks at runtime, (2) executing unstructured sparsity patterns with high utilization, and (3) supporting dual-precision computation without additional dedicated high-precision multipliers.

\subsection{Overall Architecture}
\label{sec:sec4.1}

As illustrated in Figure~\ref{fig:fig8}, APT comprises a sparsity-aware dual-precision matrix processing unit (SD-MPU), a vector processing unit (VPU), a quantizer, on-chip memory, a top controller, and a DMA module.
The SD-MPU handles $QK^\top$ and $PV$ operations via four components: (1) a mask management unit for generating, compressing, and reconstructing precision-aware masks, (2) an address translator unit for resolving irregular memory access patterns and avoiding bank conflicts, (3) 16 dual-precision MAC lines (DP-MACs) supporting 12-bit and 6-bit execution, and (4) a transpose controller for flexible operand mapping.
The VPU executes Softmax, normalization, and other non-matrix operations in 16-bit fixed-point to ensure numerical stability.
The quantizer performs symmetric uniform quantization with per-tile scaling factors stored in 16-bit integer format.
On-chip memory stores $Q$, $K$, $V$ tensors, intermediate activations, weights, and execution metadata including masks and address mappings.
The DMA's data aligner packs 6-bit operands into byte-aligned streams for efficient DRAM transactions, minimizing bandwidth waste from non-byte-aligned data.

\subsection{System-Level Dataflow}
\label{sec:sec4.2}

\begin{figure}[t]
    \centering
    \includegraphics[width=0.45\textwidth]{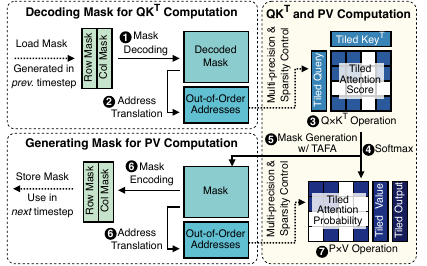}
    \vspace{-0.13in}
    \caption{Tile-based execution flow of APT}
    \vspace{-0.13in}
    \label{fig:fig9}
\end{figure}

\begin{figure}[t]
    \centering
    \includegraphics[width=0.44\textwidth]{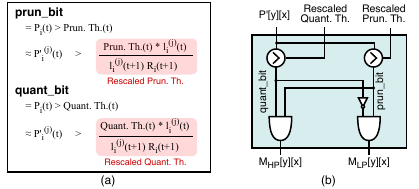}
    \vspace{-0.14in}
    \caption{(a) Hardware-level threshold computation in TAFA (b) Design of the mask generator}
    \vspace{-0.15in}
    \label{fig:fig10}
\end{figure}

Figure~\ref{fig:fig9} illustrates the system-level dataflow.
APT builds on FA's tile-based execution, incorporating sparsity-aware and dual-precision processing guided by attention probabilities.
By decoupling computation across tiles, APT eliminates global dependencies, enabling each tile to be processed independently.
While the tile serves as the scheduling unit inherited from FA, APDT applies pruning and precision decisions at element-wise granularity within each tile, assigning each element to pruned, 6-bit, or 12-bit execution.

The execution proceeds in seven stages:
\inversecircledsmall{1} High- and low-precision execution masks from the previous timestep are decoded per tile.
\inversecircledsmall{2} The address translator extracts valid element addresses into an address translation table (ATT) to guide sparse data access and avoid bank conflicts.
\inversecircledsmall{3} Tile-based $QK^\top$ is performed under sparsity and dual-precision control.
\inversecircledsmall{4} Partial Softmax generates unnormalized attention probabilities.
\inversecircledsmall{5} TAFA provides rescaled thresholds, and APDT produces new execution masks.
\inversecircledsmall{6} Masks are compressed, and a new ATT is built for $PV$.
\inversecircledsmall{7} Tile-based $PV$ is performed with input sparsity and dual-precision.
Masks are stored in external memory for reuse in the next timestep.
All mask management and address translation are overlapped with computation, avoiding additional pipeline stalls.

\subsection{Mask Management Unit}
\label{sec:sec4.3}

The mask management unit generates, compresses, and reconstructs execution masks based on TAFA-predicted attention probabilities.
Figure~\ref{fig:fig10} illustrates the mask generator and threshold rescaling.
To avoid full attention probability reconstruction, the generator rescales dual thresholds using the TAFA factor and applies them directly to $\mathbf{P}_i^{\prime(j)}(t)$, as shown in Figure~\ref{fig:fig10}(a).
This produces \texttt{prun\_bit} and \texttt{quant\_bit} signals: an element is pruned when \texttt{prun\_bit}~$=$~0; otherwise, \texttt{quant\_bit} determines low-bit (0) or high-bit (1) execution.
These signals construct high- and low-precision execution masks ($M_\text{HP}$, $M_\text{LP}$), as shown in Figure~\ref{fig:fig10}(b).

The mask encoder compresses tile-wise masks via column-wise and row-wise paths.
Each path flags rows or columns whose valid element count exceeds a threshold, producing compressed column masks (CM) and row masks (RM) applied independently to both precision masks.
Since the decoder reconstructs tile masks by applying a row-wise and column-wise OR over CMs and RMs, the mask memory footprint is reduced from $T^2$ to $2T$ per tile.
This reconstruction is conservative---it may retain elements that would otherwise be pruned, executing them at low precision rather than skipping them---and this effect is reflected in the accuracy results of Table~\ref{tbl:tbl2}.
During reconstruction, the high-precision mask is prioritized, and remaining elements are assigned to low-precision.
The refresh mechanism resets pruning masks by toggling a single control bit in the decoder, incurring negligible overhead.

\subsection{Address Translator Unit}
\label{sec:sec4.4}

Element-wise sparsity produces highly irregular memory addresses, risking bank conflicts that degrade utilization.
The address translator scans masks row by row and pushes valid addresses into per-bank FIFOs.
By popping one address per FIFO each cycle, all accesses target different banks, eliminating conflicts.
Only upper address bits are stored, excluding the lower 4 bits that indicate the target bank.
The translated addresses form the ATT, which guides sparse, precision-aware memory accesses during both $QK^\top$ and $PV$ operations, maintaining high throughput in the compute pipeline.

\begin{figure}[t]
    \centering
    \includegraphics[width=0.44\textwidth]{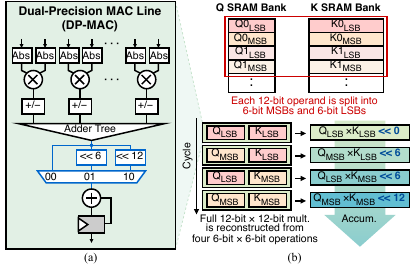}
    \vspace{-0.12in}
    \caption{(a) Architecture of DP-MAC (b) Computation flow of dual-precision}
    \vspace{-0.1in}
    \label{fig:fig13}
\end{figure}

\subsection{Dual-Precision MAC and Sparse Execution}
\label{sec:sec4.5}

Figure~\ref{fig:fig13}(a) illustrates the DP-MAC architecture.
Since APDT assigns the majority of elements ($>$50\%) to pruning and $\sim$30\% to 6-bit execution (Table~\ref{tbl:tbl2}), we design the DP-MAC to natively operate on 6-bit arithmetic, minimizing area for the dominant operation.
For the $\sim$10\% of elements requiring 12-bit precision, the DP-MAC reconstructs full-precision results by decomposing each 12-bit operand into 6-bit MSB and LSB segments stored in interleaved format, avoiding the cost of dedicated high-precision multipliers.
For 6-bit computation, only the MSB segment is accessed.
Two execution masks ($M_\text{LP}$, $M_\text{HP}$) schedule 6-bit and 12-bit operations sequentially to avoid datapath conflicts.
As shown in Figure~\ref{fig:fig13}(b), 12-bit operations are performed by sequentially processing partial products of MSB and LSB combinations, with appropriate sign correction and shifting to reconstruct the correct result.
This design enables efficient dual-precision computation without additional datapaths.

\begin{figure}[t]
    \centering
    \includegraphics[width=0.44\textwidth]{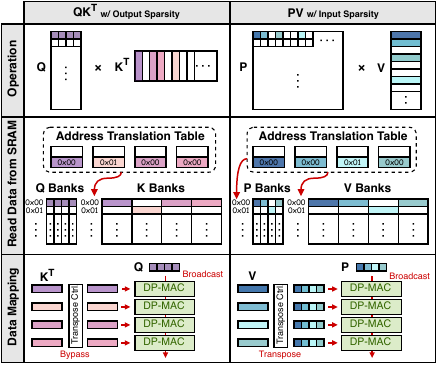}
    \vspace{-0.12in}
    \caption{Computation flow of sparse $QK^\top$ and $PV$ operations}
    \vspace{-0.15in}
    \label{fig:fig14}
\end{figure}

The SD-MPU employs 16 DP-MACs, each with a dedicated bank, performing 256 multiplications per cycle.
Figure~\ref{fig:fig14} illustrates the sparsity-aware computation flow (shown with 4 DP-MACs for simplicity).
In the $QK^\top$ operation, the query matrix $Q$ is column-wise interleaved across 16 Q banks, while $K^\top$ is organized into column blocks also interleaved across banks.
During execution, the selected row from $Q$ is broadcast to all DP-MACs, and ATT provides only valid column block addresses of $K$, exploiting output sparsity.
In the $PV$ operation, the attention probability matrix $P$ is stored column-wise, and the value matrix $V$ is partitioned into row blocks based on the sparsity pattern.
ATT provides addresses of valid $P$ and $V$ elements, enabling input sparsity.
The transpose controller realigns $V$ before sending it to the DP-MACs to resolve misaligned access directions.
Although the actual computation is sparse, ATT dynamically handles sparsity by providing only the addresses of valid data, which are densely stored in the SRAM banks.
The SD-MPU first performs sparse 6-bit execution, then selectively applies 12-bit execution, achieving efficient sparse dual-precision computation.

\section{Evaluation}
\label{sec:sec5}

\subsection{Experimental Setup}
\label{sec:sec5.1}

\begin{table}[t]
    \centering
    \caption{Hardware specifications}
    \label{tbl:tbl1}
    \vspace{-0.13in}
    \includegraphics[width=0.46\textwidth]{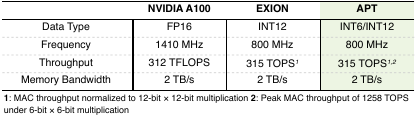}
    \vspace{-0.12in}
\end{table}

We evaluate APT on PixArt-$\alpha$~\cite{chen2023pixartalpha}, Stable Diffusion 3~\cite{10.5555/3692070.3692573}, and FLUX.1-dev~\cite{flux2024} at 1K, 2K, and 4K resolutions with batch size 1.
Accuracy is measured on COCO 2017~\cite{lin2015coco} without fine-tuning or retraining.
Per-head standard deviation statistics $\sigma_h(t)$ come from a per-model offline profiling step.
We implement APT at the RTL level in SystemVerilog and synthesize using Synopsys Design Compiler with a 14\,nm process technology.
APT operates at 800\,MHz and integrates an HBM2E memory system with 2\,TB/s bandwidth, reproducing conditions similar to the NVIDIA A100 environment.
The energy consumption of HBM2E is modeled based on DRAM vendor specifications~\cite{10413890}.
We develop a cycle-level simulator incorporating both computation and memory access latency and power.

We compare APT with NVIDIA A100, EXION~\cite{exion}, and APT-Base (same hardware, no pruning/quantization).
EXION is reimplemented following the original paper's specifications.
A100 executes FP16 inference with FlashAttention~\cite{dao2023flashattention2}; latency is measured via CUDA events and power via NVIDIA-SMI.
All configurations are scaled to match A100's peak throughput (Table~\ref{tbl:tbl1}).
APT uses 3,072 SD-MPUs and a 16\,MiB buffer; EXION uses 48 cores and a 128\,MiB buffer.
APT executes all DiT block operations internally, with all weights and activations residing in HBM2E, requiring host-device transfer only at initialization and final output.

\begin{table}[t]
    \centering
    \caption{Model configurations and accuracy evaluation}
    \vspace{-0.13in}
    \label{tbl:tbl2}
    \includegraphics[width=0.46\textwidth]{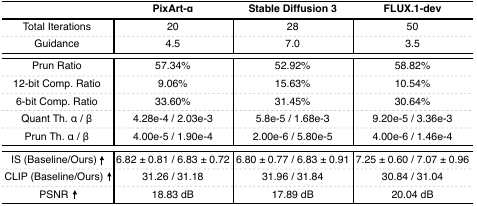}
    \vspace{-0.115in}
\end{table}

\subsection{Accuracy Evaluation}
\label{sec:sec5.2}

Table~\ref{tbl:tbl2} summarizes accuracy results using inception score (IS)~\cite{salimans2016isscore}, CLIP score~\cite{hessel2022clipscor}, and PSNR, evaluated at 1K with prompts and seeds separate from profiling.
Across all metrics, APT maintains accuracy comparable to the FP16 baseline.
IS and CLIP variations remain within 2.5\%, and PSNR indicates negligible perceptual loss.
PixArt-$\alpha$ preserves FP16-level quality with 57.34\% pruning and only 9.06\% high-precision computation.
SD3 and FLUX show similar trends despite their different architectural designs, confirming that APDT generalizes across DiT variants.
These results confirm that attention probabilities serve as an effective unified importance metric for guiding element selection and precision assignment across diverse diffusion transformer architectures.

\begin{figure}[t]
    \centering
    \includegraphics[width=0.445\textwidth]{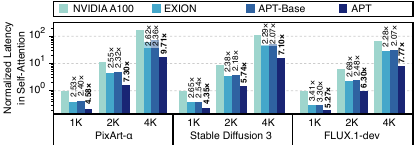}
    \vspace{-0.13in}
    \caption{Comparison of self-attention latency}
    \vspace{-0.115in}
    \label{fig:fig16}
\end{figure}

\begin{figure}[t]
    \centering
    \includegraphics[width=0.445\textwidth]{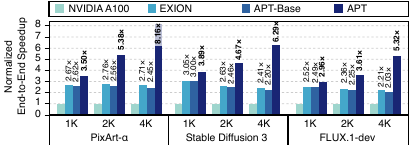}
    \vspace{-0.13in}
    \caption{Comparison of end-to-end latency speedup}
    \vspace{-0.15in}
    \label{fig:fig19}
\end{figure}

\subsection{Performance Evaluation}
\label{sec:sec5.3}

\subsubsection{Self-Attention Latency}
\label{sec:sec5.3.0}

Figure~\ref{fig:fig16} presents the normalized self-attention latency.
APT achieves $4.35\times$--$5.27\times$ speedup over A100 and $1.55\times$--$1.81\times$ over EXION at 1K, reaching up to $9.71\times$ and $3.70\times$ at 4K as attention increasingly dominates.
APT-Base (same hardware without APDT/TAFA) already outperforms A100 due to its optimized tile-based dataflow, demonstrating that the architectural design provides a strong baseline even without algorithmic optimizations.
The full APT further improves upon APT-Base by combining APDT and TAFA to drive aggressive pruning and precision selection, with the gap between APT and APT-Base widening at higher resolutions as increased sparsity amplifies the benefit of attention probability--guided optimization.

\subsubsection{End-to-End Performance}
\label{sec:sec5.3.1}

Figure~\ref{fig:fig19} shows end-to-end speedup across models and resolutions.
APT achieves up to 3.89$\times$ over A100 at 1K, growing to 8.16$\times$ at 4K, and outperforms EXION by up to 3.01$\times$.
Since self-attention dominates end-to-end latency at high resolutions (Figure~\ref{fig:fig1}), APT's attention-level acceleration directly translates into end-to-end improvement.
A100 suffers from low utilization due to kernel launch and synchronization overheads, 
while EXION realizes only about 20\% sparsity in self-attention and 3\%--5\% in feed-forward layers on modern DiTs, limiting its speedup.
Cross-attention contributes only a minor portion of total computation due to its short, resolution-independent token length, and is executed in dense full-precision mode.
The gap between APT and existing platforms widens at higher resolutions, confirming the scalability of attention probability--guided optimization.

\begin{figure}[t]
    \centering
    \includegraphics[width=0.445\textwidth]{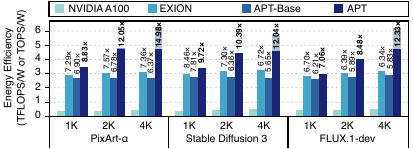}
    \vspace{-0.13in}
    \caption{Comparison of energy efficiency}
    \vspace{-0.12in}
    \label{fig:fig20}
\end{figure}

\begin{table}[t]
    \centering
    \caption{Breakdown of area and power usage}
    \vspace{-0.13in}
    \label{tbl:tbl3}
    \includegraphics[width=0.46\textwidth]{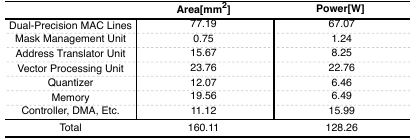}
    \vspace{-0.12in}
\end{table}

\subsubsection{Energy Efficiency and Area/Power Breakdown}
\label{sec:sec5.3.2}

Figure~\ref{fig:fig20} compares energy efficiency, measured as effective throughput per watt across the full denoising pipeline.
APT delivers up to 14.98$\times$ higher efficiency than A100 and 2.04$\times$ higher than EXION across all models and resolutions.
APT achieves this by pruning low-importance elements and assigning dual-precision via attention probabilities, while maintaining high compute utilization through its tile-based architecture.

Table~\ref{tbl:tbl3} provides the area and power breakdown.
The total chip area is 160.11\,mm$^2$ and power is 128.26\,W at 0.8\,V/800\,MHz.
DP-MACs dominate at 48.21\% area and 52.29\% power (1.36\,mW per DP-MAC).
The mask management and address translator units enabling APDT and TAFA account for 10.26\% of area and 7.40\% of power, demonstrating that the co-design overhead for runtime dynamic sparsity and dual-precision control is modest.

\begin{figure}[t]
    \centering
    \includegraphics[width=0.445\textwidth]{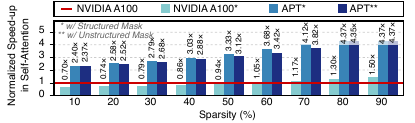}
    \vspace{-0.15in}
    \caption{Self-attention speedup under varying sparsity}
    \vspace{-0.15in}
    \label{fig:fig21}
\end{figure}

\subsection{Comparison with GPU-Based Optimization}
\label{sec:sec5.4}

GPUs face inherent limitations for exploiting sparsity in DiT self-attention.
Figure~\ref{fig:fig21} evaluates GPU sparsity performance using an optimized sparse attention kernel~\cite{guo2024blocksparse} on A100 at 1K.
Even under structured sparsity---more GPU-friendly than APDT's element-wise pattern---A100 remains slower than the dense FlashAttention execution up to 50\% sparsity, due to kernel launch overhead, synchronization, and scattered memory accesses.
Since APDT induces unstructured sparsity at runtime, GPU performance is expected to degrade further under such irregular access patterns.
Low-bit quantization on GPUs is also limited: as noted in Section~\ref{sec:sec2.3}, SVDQuant~\cite{li2024svdquant} leaves $QK^\top$ and $PV$ in full precision, and since these operations dominate computation at high resolutions, GPU-side quantization cannot address the attention bottleneck that APT directly targets.

\begin{figure}[t]
    \centering
    \includegraphics[width=0.445\textwidth]{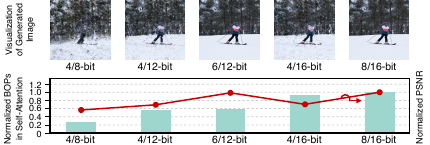}
    \vspace{-0.11in}
    \caption{Comparison of dual-precision configurations}
    \vspace{-0.1in}
    \label{fig:fig22}
\end{figure}

\begin{figure}[t]
    \centering
    \includegraphics[width=0.445\textwidth]{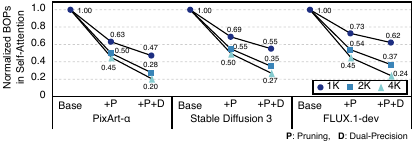}
    \vspace{-0.15in}
    \caption{Analysis of computation reduction in APT}
    \vspace{-0.15in}
    \label{fig:fig17}
\end{figure}

\subsection{Design Space Analysis}
\label{sec:sec5.5}

\subsubsection{Dual-Precision Configuration}
\label{sec:sec5.5.1}

Figure~\ref{fig:fig22} evaluates several dual-precision configurations.
We extend the BOPs metric~\cite{10.1145/3444943} to dual-precision, counting bit-operations per multiplication.
The 6/12-bit configuration yields lower BOPs than 4/16-bit despite the latter's lower base precision.
Configurations involving 4-bit computation introduce visible artifacts and a clear PSNR drop.
The 6/12-bit achieves perceptual quality nearly identical to 8/16-bit with fewer bit-operations, providing the best quality--computation trade-off.
This analysis directly informed our DP-MAC design choice of 6-bit native arithmetic with 12-bit reconstruction (Section~\ref{sec:sec4.5}).

\subsubsection{Computational Efficiency}
\label{sec:sec5.5.2}

Figure~\ref{fig:fig17} shows the BOPs reduction under three configurations: APT-Base (Base), APT with pruning only (+P), and APT with both pruning and dual-precision (+P+D).
Pruning alone provides substantial reduction; dual-precision amplifies it further.
At 4K, pruning reduces BOPs by 53.7\% on average, and combining both reaches up to 76.4\%.
Both techniques become increasingly effective as resolution grows, since self-attention's share of total computation increases with resolution (Figure~\ref{fig:fig1}), amplifying the benefit of sparsity and reduced precision.

\section{Related Works}
\label{sec:sec6}

Recent diffusion accelerators~\cite{ditto, tang2025diff, exion} leverage temporal similarity but face limitations.
Ditto~\cite{ditto} and Diff-DiT~\cite{tang2025diff} leverage inter-timestep activation differences for low-bit computation, but both are limited to at most 8-bit precision, leading to noticeable quality degradation on modern DiTs.
Moreover, Ditto must recover full activations from off-chip memory for attention, incurring memory overhead, while Diff-DiT's approximate attention is only applicable to a subset of timesteps and its effectiveness diminishes with fewer steps.
EXION~\cite{exion} combines intra-timestep top-$k$ pruning in attention with inter-timestep approximation in feed-forward layers, but achieves only limited sparsity on modern DiTs and handles only output sparsity.
In contrast, APT predicts attention probabilities to guide element-wise pruning and dual-precision decisions, and its architecture supports both input and output sparsity.

Attention accelerators such as Sanger~\cite{lu2021sanger}, FACT~\cite{fact}, and SpAtten~\cite{wang2021spatten} require explicit attention score construction, which is incompatible with FA's tiled execution.
Moreover, these methods target language models and do not account for evolving attention distributions across diffusion timesteps.
APT avoids these limitations through TAFA, which predicts attention probabilities from the previous timestep's normalization statistics within FA's tiled execution, requiring no additional score prediction stage.

Sparse quantization techniques~\cite{jeong2024sdq, dettmers2023spqr} allocate precision based on static weight statistics, targeting weight--activation linear operations.
In contrast, DiT attention involves activation--activation operations whose importance evolves across timesteps, which APT captures through attention probabilities as a dynamic importance signal.
Structured sparse attention methods~\cite{jiang2024minference, yuan2025native, yuan2024ditfastattn, zhang2025spargeattention} reduce attention cost at block or token-group granularity, missing fine-grained importance variations.
APT operates at per-element granularity and is complementary to these structured approaches.

Software-level techniques such as ToCa~\cite{zou2024accelerating} and DuCa~\cite{zou2024DuCa} cache features across timesteps to reduce redundant computation.
However, ToCa's full attention map construction is incompatible with FA, while DuCa achieves only limited speedup.
APT combines algorithmic optimization with co-designed hardware, integrating APDT's decisions with the SD-MPU's mask management and address translation to handle runtime dynamic sparsity.
\section{Conclusion}
\label{sec:sec7}

This paper presents APT, a software--hardware co-designed accelerator for high-resolution DiT models that addresses the self-attention bottleneck through attention probability--guided pruning and quantization.
At the algorithm level, APDT dynamically prunes and quantizes attention elements based on their attention probabilities, substantially reducing computation while preserving output quality across three SOTA DiT models.
TAFA enables APDT within FlashAttention by reusing normalization statistics across timesteps, avoiding full probability materialization.
At the architecture level, the SD-MPU's mask management, address translation, and dual-precision MAC units are co-designed with APDT to efficiently handle the resulting element-wise dynamic sparsity with modest area overhead.
APT achieves up to 8.16$\times$ and 3.01$\times$ speedup, and up to 14.98$\times$ and 2.04$\times$ higher energy efficiency over A100 and EXION, respectively, with the advantage growing at higher resolutions.

\begin{acks}
This work was partly supported by Institute for Information \& communications Technology Promotion(IITP) grant funded by the Korea government(MSIT) (No. RS-2025-02264029, Integration and Validation of an AI Semiconductor-Based Data Center Training and Inference System) and by Samsung Electronics Co., Ltd(IO251211-14291-01).
\end{acks}

\bibliographystyle{ACM-Reference-Format}
\clearpage
\bibliography{refs}

\end{document}